\documentclass[twocolumn,showpacs,preprintnumbers,amsmath,amssymb]{revtex4}

\usepackage{graphicx}
\usepackage{dcolumn}
\usepackage{bm}
\usepackage{epsfig}

\newcommand \bea{\begin{eqnarray}}
\newcommand \eea{\end{eqnarray}}

\newcommand \la{\raisebox{-.5ex}{$\stackrel{<}{\sim}$}}
\newcommand{\av}[1]{\langle{#1}\rangle}

\begin{document}


\title{Extended Bose-Hubbard model with 
incompressible states at fractional numbers}

\author{H. Heiselberg}
\altaffiliation[Also at ]{Univ. of S.Denmark}
\email{hh@ddre.dk}
\affiliation{Danish Defense Research Establishment, Ryvangsalle' 1, 
DK-2100 Copenhagen \O, Denmark}


\begin{abstract}

The Bose-Hubbard model is extended to include nearest and far neighbor
interactions and is related to the fractional quantum Hall effect
(FQHE). Both models may be studied in optical lattices with quantum
gases.  The ground state is calculated for the extended Bose-Hubbard
model with strong repulsive interactions (weak
hopping). Incompressible Mott insulator states are found at rational
filling fractions compatible with the principal and secondary FQHE
filling fractions of the lowest Landau levels observed experimentally.
It is discussed to which extent these states at fractional filling
survive or undergoes a Mott insulator transition to a superfluid as
hopping terms are included.

\end{abstract}

\maketitle

\section{Introduction}

Quantum gases provide new insight in strongly correlated quantum
systems and phase transitions in systems of interacting
bosons and fermions. A variety of interesting phases exist in various
dimensions such as superfluids, Mott insulators, charge density waves,
incompressible states
in the integer and fractional quantum Hall effect (FQHE)
\cite{Klitzing,Tsui,Pan,Laughlin,Prange}, etc.  Lattices appear
naturally in solids and spin models \cite{Ising,Bethe,Stephenson,Maj},
and recently also optical lattices with Bose atoms have been studied
\cite{Fischer,Zwerger}. The superfluid to Mott insulator transition
has been observed \cite{Greiner} as predicted from the Bose-Hubbard
model. It has been suggested \cite{Sorensen} that by applying an
additional magnetic field the optical lattices provide a clean system
without impurities, where the FQHE can be studied in
detail. Alternatively, rapidly rotated harmonic oscillator traps with
cold Fermi or Bose atoms may also form a 2D Hall fluid
\cite{Wilkin,Ho,Paredes}.

These possibilities stimulate renewed interest in the long standing
problem of the FQHE and its numerous incompressible quantum liquid
states at mysterious filling fractions of the lowest Landau levels
\cite{Klitzing,Tsui}. In a recent experiment \cite{Pan} new FQHE
fractions are found at 3/8, 4/11, 5/13, 6/17, 7/11, etc., which fall
outside the Laughlin fractions $1/(2m+1)$ and Jain fractions
$m/(2m\pm1)$ for integer $m$.

The Hubbard model is widely used to describe a number of systems such as
strongly correlated electrons. 
Quantum gases of Bose atoms in optical lattices have recently been
described by the Bose-Hubbard model, which in one dimension reads
\bea \label{BH}
  H_{BH}= \sum_i t_1(b_i^\dagger b_{i+1} + b_{i+1}^\dagger b_i) \,+\,
  U_0 n_i (n_{i}-1) \,.
\eea
Here, $t_1$ is the hopping parameter to nearest site and $U_0$ is the
on-site interaction energy. The system is a superfluid except in the
cases where the filling fraction is precisely an integer and the
hopping term is sufficiently small, where a transition occurs from a
superfluid to a Mott insulator as observed in \cite{Greiner}.
Including nearest neighbor interaction leads to charge density waves
and crystalline structure at half filling and a related Mott transition
\cite{Scalettar,Monien,Paredes}.

We shall be interested in extended Bose-Hubbard models (EBHM)
including not only on-site interactions $U_0$ but also interactions
$U_n$ between particles $n=1,2,3,4,...$ sites away.  In the strongly
interacting (atomic) limit or equivalently weak hopping, $t_1\ll U_n$, the
extended Bose and Fermi Hubbard models reduce to well known spin
models. The Ising model \cite{Ising} has only nearest neighbor $U_1$
interactions but next-nearest neighbor interactions $U_2$ have also been
included (see, e.g., \cite{Stephenson,Maj}).

The purpose of this work is to point to a connection between the
Hamiltonian for the FQHE and the EBHM as outlined in section II.
Subsequently in section III to find the ground state of the EBHM in
the strongly interacting limit, calculate its energy, chemical
potential, pressure, etc. Most importantly, a spectrum of
incompressible Mott insulator states are found whenever the filling
fraction is a rational number $q/n$. In section IV the Bose filling
fractions are related to the fermion FQHE case, and it shown that they
contain the Laughlin and Jain sequences \cite{Jain,Tsui} as well as
all the secondary FQHE fractions found experimentally
\cite{Pan}. Finally in section V the effect of hopping terms and the
survival of Mott insulator states at fractional filling is discussed.

\section{Approximating the FQHE as an extended Bose-Hubbard model}

The fractional quantum Hall systems \cite{Prange} can be essentially
reduced to the problem of $N$ spinless electrons with repulsive
interactions in $M$ available states \cite{Dyakonov}. In the disk
geometry the states are $\psi_k(z)= z^k\exp(-|z|^2/4)/\sqrt{2\pi 2^kk!}$,
$k=0,1,...,M-1$; $z=x+iy=re^{i\phi}$ in units of the magnetic length $l$
\cite{area}.

When the filling fraction is exactly $\nu=N/M=1$, 1/3, 1/5, 1/7, etc.,
 the ground state is the antisymmetric Laughlin wave function 
\bea \label{Psi}
  \Psi_{m} = A
  \prod_{i<j}^{N}(z_i-z_j)^{m} \exp\left(-\sum_k^N|z_k|^2/4\right)
  \,,
\eea
with $m=1$, 3, 5, 7, etc., and normalization factor $A^2$.
Analogously, a symmetric Laughlin wave function with
$m=2,4,6,...$ can be separated for bosons instead of fermions. 

In the lowest Landau level ($N\le M$) the total wave function also
contains a symmetric part $\Psi_S$
\bea
  \Psi =  \Psi_S(z_1,...,z_N) \Psi_{m} \,.
\eea
The remaining FQHE problem is to find the incompressible states of $\Psi_S$
besides the fractions $\nu=1,1/3,1/5,...$
implicit in the Laughlin wave function. At these fractions $\Psi_S=1$. 

In the following we shall concentrate on the lowest Landau level,
$N\le M$, and study the symmetric wave function $\Psi_S$ for $N$
particles in $M_B=M-mN$ available states. Here the number of states,
that the Laughlin wave function use up, have been subtracted.  By this
``bosonization'' the N particles has thus become N spinless bosons in
$M_B$ states with a symmetric many-boson wave function $\Psi_S$ (see, e.g.
\cite{MacDonald} for details on bosonization). 
The Bose filling fraction $\nu_B=N/M_B$ is related
to the fermion one by
\bea\label{nu}
  \nu = \frac{N}{M} = \frac{\nu_B}{1+m\nu_B}  \,,
\eea
with $m=1$, 3, 5,....

We can choose any other orthogonal basis from $\psi_k(z)$. This may be
convenient if a many-body wave function can be constructed from this
new basis such that the Hamiltonian approximately diagonalizes.
This would greatly simplify the search for the ground state of the FQHE.
A candidate for such a convenient basis are 
the localized Wannier-type wave functions
\bea \label{Phi}
  \Phi_j(z) &=&  \frac{1}{\sqrt{2\pi M_B}}
      \sum_{k=0}^{M_B-1} \psi_k(z)\, e^{-i2\pi kj/M_B} \,.
\eea
where $j=0,1,...,M_B-1$. 
As described in Appendix A these wave functions are localized in angle 
$\phi$ around $2\pi j/M_B$.

Dyakonov argues in Ref. \cite{Dyakonov} the similar localized
wave functions are useful for constructing a crystal-like many-body
state that contains at least some of the right features of the FQHE
ground state.  The localized wave functions of Ref. \cite{Dyakonov} are
an approximate model for the FQHE on a circle. It is simple and
illustrative, and is therefore described in Appendix A.

It should be emphasized that the localized states are simply another
basis. They are not generated by an external periodic potential as,
e.g., an atomic lattice. Whether the basis is ``better'' will depend
on whether the repulsion between particles dominates over hopping
leading to localized vs. conduction states respectively. It will be
argued below that localized states of Eq. (\ref{Phi}) may be the
preferred ones.

The Hamiltonian in the lowest Landau level 
is now a one dimensional lattice model
containing only repulsive two-body interaction $V$ between the particles
\bea \label{HV}
  H = \sum_{1\le i<j\le N}  V(z_i-z_j) \,.
\eea
The interactions can in the following be almost any repulsive potential
including the standard Coulomb: $V=e^2/|z_1-z_2|$.
The delta function potential should be treated with care as discussed in
Appendix A.

The kinetic energies and other degrees of freedom are absorbed in the
cyclotron energies, $\hbar\omega_c=eB/mc$, of the lowest Landau
level. The cyclotron energies 
have been subtracted so that only interaction energies
remain in the Hamiltonian. In this sense the system is strongly
interacting.  All energies of this Hamiltonian will scale with the
overall strength of the two-body potential.

In tight-binding models the interaction is often written
in second quantized form as 
\bea \label{H}
   H =\frac{1}{2} \sum_{ij,i'j'} V_{ij,i'j'} 
          b^\dagger_i b^\dagger_j b_{i'} b_{j'}  \,,
\eea
where $b_i^\dagger$ is the usual creation operator of a state
$\Phi_i$ at site $i$, and  
\bea \label{Vij}
  V_{ij,i'j'} &=& \int \Phi_i(z_1)\Phi_j(z_2) 
                  \Phi^*_{i'}(z_1)\Phi^*_{j'}(z_2) |\Psi_m(z_1,z_2,....)|^2
   \nonumber \\
    &&  V(z_1-z_2)dz_1 dz_2   \,,
\eea
are the four-center integrals. Here, the Laughlin wave function present in
the total wave function is included in which
all other particles have been integrated out. $|\Psi_m|^2$ does not affect
the integrals much except that it contains a correlation hole 
$\sim (z_1-z_2)^{2m}$ which screens short distances of the two-body 
interaction.

The largest of the $V_{ij,i'j'}$ are the interaction between
particles on just two sites, the direct $i=i',j=j'$ and the exchange
 $i=j',j=i'$. 
As described in Appendix A the other overlap integrals are small because
the localized wave functions are rapidly oscillation in angle and interfere
destructively.
We therefore extract the largest (tight binding) direct and exchange part of
the FQHE Hamiltonian of Eq. (\ref{H})
\bea \label{H0}
   H_0 &=&\frac{1}{2} \sum_{ij,ij} (V_{ij,ij} + V_{ij,ji}) 
   n_i [n_j-\delta_{i,j}]   \,.
\eea
Here, $n_i=b^\dagger_i b_i$ is the number operator on site $i$.

For long range interactions such that the two-body interaction can
be treated as a constant, $V\simeq V_0$, only the direct interaction
survives, i.e. $V_{ij,i'j'}=V_0\delta_{i,i'}\delta_{j,j'}$, and 
therefore $H=H_0$. This is the mean field
limit with energy $E=V_0M_B \nu_B^2$.

For finite range interactions terms will remain in the Hamiltonian
that connect different orbitals. These will be denoted
\bea
  T = H-H_0   \,.
\eea
$T$ contains hopping
terms between nearest neighbors as the Hubbard Hamiltonian but also
contains hopping terms to any neighbor state $n$. As the wave functions are
rapidly oscillating as described in Appendix A and not in phase in $T$, their
overlap is generally smaller as those in $H_0$. 
We shall in section V discuss the effect of the
hopping terms in $T$ and possible Mott-Hubbard transitions if $T$ is strong.

As the direct and exchange potentials depend only on the distance
$n=j-i$ between the two interacting particle sites, 
we can rewrite the couplings in $H_0$ as
\bea \label{J}
   U_n &=& \frac{1}{2}\left(V_{ij,ij} + V_{ij,ji}\right)   \,.
\eea
Note that $U_{M_B-n}=U_n$. 
In the following this double counting is
avoided by truncating $n\le M_B/2$ and multiplying by a factor of
two.

In the following only $U_n$ will be important whereas
the details of the localized wave function and the two-body interaction 
are contained in $U_n$.
We shall only assume that the two-body interactions
are repulsive and decreasing with range as is the case for Coulomb
repulsion between electrons.  As result the resulting couplings $U_n$
will also be repulsive and decreasing functions of the neighbor
distance, i.e. $U_n>U_{n+1}\ge0$.

Fortunately, as will be shown in the following section, 
the Hamiltonian $H_0$ can be solved for
rather general couplings $U_n$ and a number of qualitative results can be
drawn such as incompressible states at fractional fillings.
Whether $H_0$ is a good approximation for the FQHE Hamiltonian $H$ as
argued above will depend on the effect of $T=H-H_0$ on the ground state
as will discussed in section V.

\section{The ground state of the EBHM in the strongly interacting limit}
 
Eqs. (\ref{H0}) and (\ref{J}) result in a
1D Bose-Hubbard Hamiltonian in the strongly interacting
limit extended with far-neighbor interactions $U_n$
but without hopping terms
\bea
  H_0= \sum_{i=0}^{M_B-1} \sum_{n\ge0}^{M_B/2} U_n n_i [n_{i+n}-\delta_{n,0}]
  \,.
\eea
The EBHM also describes optical lattices and reduces to $H_0$ for strong
lattices in the
strongly interacting limit near Feshbach resonances \cite{Fischer,Zwerger}.

The Bose-Hubbard model has only on-site interactions, i.e. only $U_0$
is non-zero. The Ising \cite{Ising} spin model ($\sigma_i=n_i-1/2$)
has only nearest neighbor interaction, i.e. only $U_1$ is non-zero. A
model including next-nearest neighbor interactions, i.e. $U_2\ne0$ was
solved by Stephenson \cite{Stephenson} and Majumdar \& Gosh \cite{Maj}.  
These statistical models mostly investigate phase
transitions and critical exponents in one or more dimensions for fixed
chemical potential.  We are here interested in the ground state (zero
temperature) in one dimension in the thermodynamic limit $N,M_B\to\infty$ 
for constant density or filling fraction
$\nu_B=N/M_B$.

We shall investigate a general class of EBHM with strong repulsive
interactions, i.e. Hamiltonians $H_0$ with repulsive $U_n\ge 0$,
$n=0,1,2,3,$ etc., that decrease with distance such that
$U_n>U_{n+1}$, i.e.  $dU_n/dn<0$. Eventually we may expect that $U_n$
becomes negligible at large $n$ as compared to other effects from impurities,
finite temperature or the experimental resolution.
In addition, we shall assume convexity \cite{Hubbard}
\bea \label{convex}
   U_{n+1}+U_{n-1}-2U_n> 0 \,,
\eea
i.e. $d^2U_n/dn^2>0$. 

The ground state is then organized as to minimize its
energy. This is achieved by distributing the particles and also
vacancies as ``evenly'' as possible avoiding ``clumping''.  We shall
argue that the particles sites (and vacancies) are ordered with a
periodicity or sequences of replicas depending on the filling
fraction (see also \cite{Hubbard}).

\subsection{Fraction of n'th neighbors, $F_n$}

As an example, consider the filling fraction 5/9. The ground state is
build up of a series of $M_B/9$ identical sequences (replicas)
repeated after each other of the following type
\bea 
  ...... \,|\, \bullet\bullet\circ\bullet\circ\bullet\circ\bullet\circ \,|\,
  \bullet\bullet\circ\bullet\circ\bullet\circ\bullet\circ  \,|\, .....
  \quad \nu_B=5/9 \nonumber
\eea
Here: $\bullet$=site with a particle and
$\circ$=hole or vacancy. The delimiter $|$ separates replicas.
Interchanging any $\bullet$ with a $\circ$ increases
the energy since $U_n>U_{n+1}$, and it is therefore the ground state.
By translating the whole lattice the nine degenerate ground states
are obtained. However, they any not connected by any two-body interactions. 

This principle of ``evenly ordering'' generalizes to any rational number for
the filling fraction $N/M_B=q/n$. i.e.  $q$ particles
in $n$ states. If $q\le n$ there will be $(n-q)$ vacancies.
The minimum energy state is when the $q$ particles (or $n-q$ vacancies)
are as distributed over the $n$ sites avoiding clumping.

\begin{figure}
\begin{center}
\psfig{file=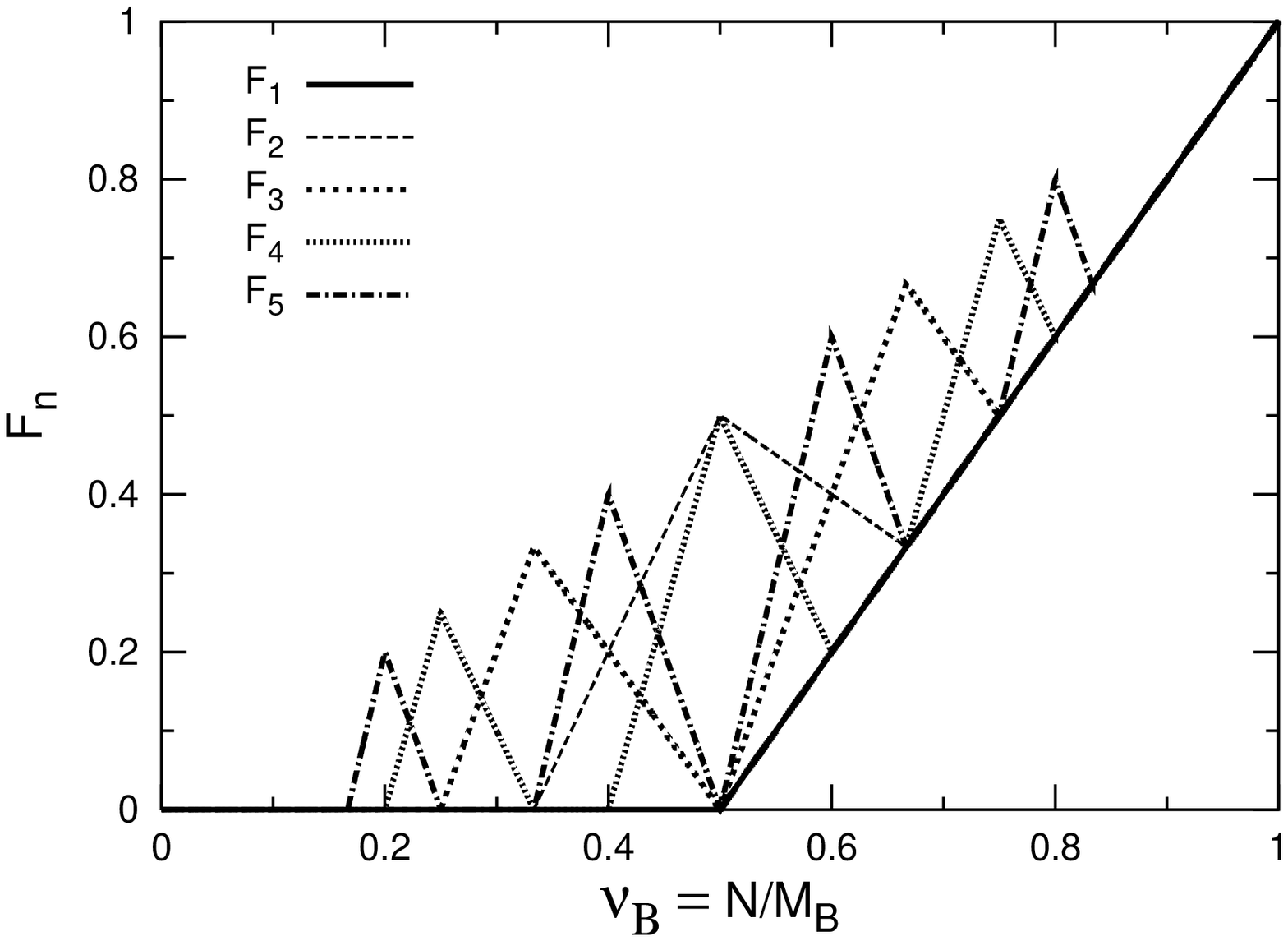,height=8.0cm,angle=-90}
\vspace{.2cm}
\begin{caption}
{The fraction of occupied sites with a neighbor $n$ sites away, $F_n(\nu_B)$,
for $n=1,2,3,4,5$.  
}
\end{caption}
\end{center}
\label{f1}
\end{figure}

\begin{figure}
\begin{center}
\psfig{file=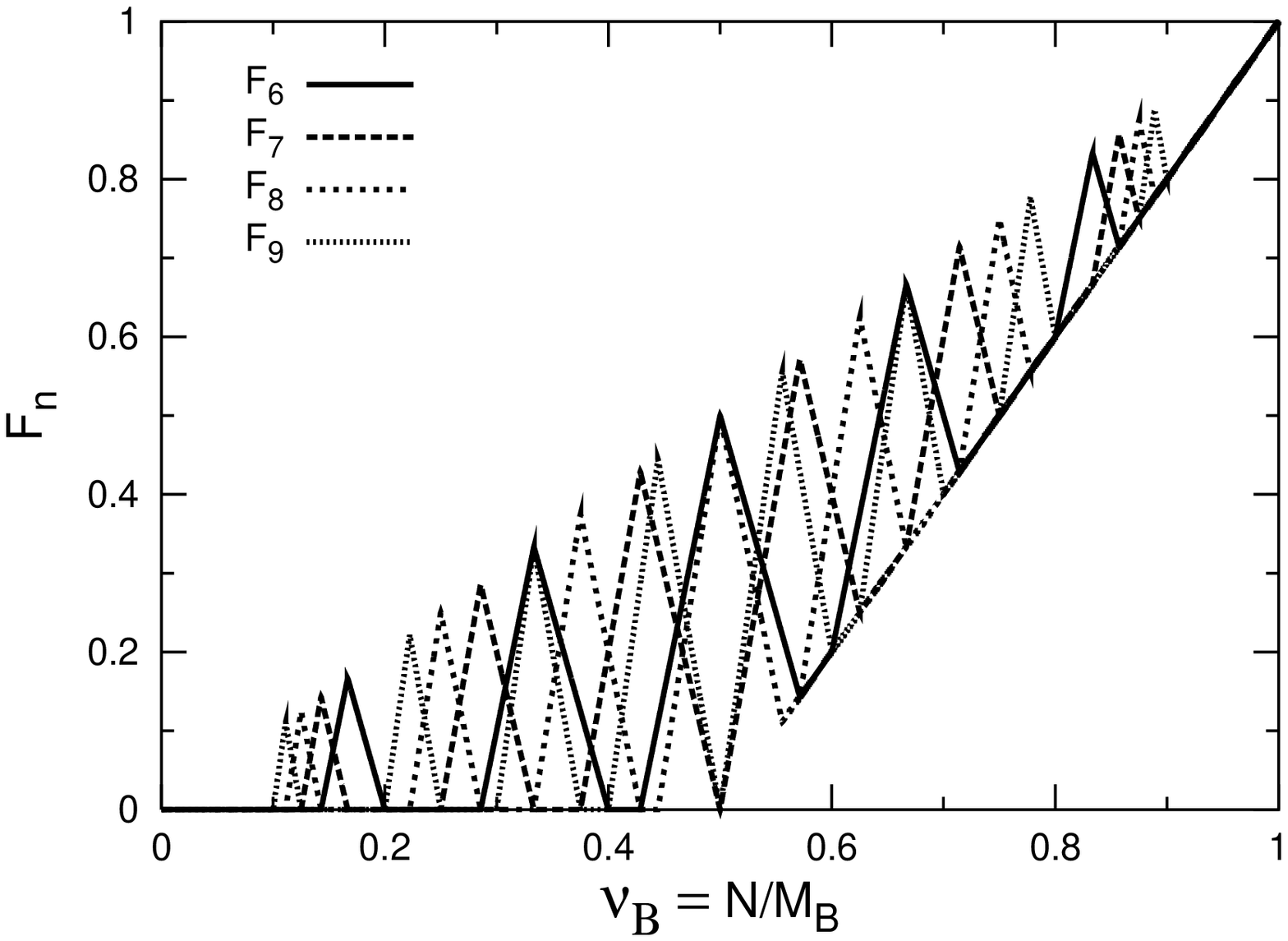,height=8.0cm,angle=-90}
\vspace{.2cm}
\begin{caption}
{As Fig. 1 but for $n=6,7,8,9$.  
}
\end{caption}
\end{center}
\label{f1b}
\end{figure}

In the thermodynamic limit $N,M_B\to\infty$ it is now a matter of combinatorics
to add up the number of $n$'th neighbors.
The ground state energy  is necessarily 
a linear function of the couplings $U_n$
\bea \label{E}
  E(N/M_B) = M_B \sum_{n=0}^{M_B/2} F_n(\nu_B) U_n \,.
\eea
The coefficients $F_n(\nu_B)$ are functions of the filling
fraction. They are the fractions of sites occupied with an occupation
also $n$ sites to one side (not to both sides in order to avoid double
counting).  In other words $M_BF_n(\nu_B)$ is the number of sites
occupied which also has a particle displaced $n$ sites to one side
(only to one side when the $n$ and $M_B-n$ double counting is avoided).
This fraction is necessarily smaller than the filling fraction,
$F_n(\nu_B)\le\nu_B$. The ratio $F_n(\nu_B)/\nu_B$ is the conditional
probability for an occupied site to have a neighbor occupation $n$
sites to one side.

\begin{table}
\caption{\label{tab:table1} The coefficients in $F_n=\alpha_n\nu_B+\beta_n$}
\begin{ruledtabular}
\begin{tabular}{||r|r|l||}
 $\alpha_n$ & $\beta_n$ & range of $\nu_B$ for $q=0,1,...,[n/2]$\\
\hline
  0 & 0 & [ q/(n-1) , (q+1)/(n+1) ] \\
  1+n & -q & [ q/(n+1) , q/n ] \\
  1 - n & q & [ q/n , q/(n-1) ]\\  \hline
 $\alpha_n$ & $\beta_n$ & range of $\nu_B$ for $q=n-1,n-2,...,[n/2]$\\
\hline
  1+n & -q & [ (q-1)/(n-1) , q/n ] \\
  1 - n & q & [ q/n , (q+1)/(n+1) ]\\
  2 & -1 & [ q/(n+1) , (q+1)/(n+1) ]\\
\end{tabular}
\end{ruledtabular}
\end{table}

It will be shown below that the functions $F_n$ 
are piece-wise linear functions of $\nu_B$ (see also Figs. 1+2) 
\bea \label{Fn}
  F_n = \alpha_n \nu_B +\beta_n   \,.
\eea
The prefactors
$\alpha_n$ and $\beta_n$ are discontinuous at the fractions but
constant between. They are in fact simple integers of 
changing signs at the fractions $\nu_B=q/n$, $q=1,2,....,n-1$.

Before deriving their general dependence on $\nu_B$ analytically a few
examples illustrates the general idea.
At half filling the state is
\bea 
  .... |\bullet\circ|\bullet\circ|\bullet\circ|\bullet\circ|
  \bullet\circ|\bullet\circ|\bullet\circ|\bullet\circ|  .....
  \quad \nu_B=1/2 \nonumber
\eea
This is the charge density wave phase of crystalline order at half integer
density. This incompressible Mott insulator state also undergoes a transition
to a superfluid at a certain hopping strength \cite{Scalettar,Monien,Paredes}.

Its energy is from Eq. (\ref{E})
\bea
  E(1/2) = N \sum_{n=2,4,6,...} U_n   \,,
\eea
i.e., $F_n=0$ for odd $n$ and $F_n=1/2$ for even $n$.

Let us next study $F_1$ and $F_2$ with increasing filling.
The fraction of occupied neighbor states,
$F_1$, is zero up to half filling, $\nu_B\la1/2$. Above it increases
linearly as $F_1=2\nu_B-1$ reaching unity at full filling $\nu_B=1$.
The fraction of next-nearest neighbors is zero for $\nu_B\la1/3$ and
then increase linearly as  $F_2=3\nu_B-1$ up to half filling,
where after it decreases linearly
as $F_2=1-\nu_B$ up to $\nu_B=2/3$.
\bea 
  ...|\bullet\bullet\circ|\bullet\bullet\circ|\bullet\bullet\circ|
  \bullet\bullet\circ|\bullet\bullet\circ|\bullet\bullet\circ|  .....
  \quad \nu_B=2/3 \nonumber
\eea
For filling above $\nu_B\ge 2/3$ the number of 
next-nearest neighbors again increase as $F_2=2\nu_B-1$ just as $F_1$.

The ordering principle implied by the condition $U_{n+1}<U_n$ allows
us to calculate the probability coefficients $F_n$ generally. Let us
start at low filling fraction.  For $\nu_B<1/(n+1)$ there are no $n$'th
neighbors and $F_n=0$.  The coefficient now increase linearly up to
$\nu_B=1/n$ where every particle has an $n$'th neighbor and therefore
$F_n=\nu_B=1/n$. By increasing the filling fraction further the number
of n'th neighbors decrease linearly and in fact vanish again at
$\nu_B=1/(n-1)$. $F_n$ now remains zero from $\nu_B=1/(n-1)$ up to
$\nu_B=2/(n+1)$ where the zig-zag behavior repeats.  At $\nu_B=2/n$
every particle again has a neighbor $n$ sites away (and also one in
between at $n/2$ for even $n$ and at $(n\pm1)/2$ for odd $n$). Therefore
$F_n(\nu_B)=\nu_B=2/n$

This zig-zag behavior of $F_n$ in fact repeats itself around each
$\nu_B=q/n$, $q=1,2,...,n-1$ (see table 1 and Appendix B) with the
same $\alpha_n$ and $\beta_n$. Above $\nu_B\ge1/2$, however, with
$F_n=2\nu_B-1$ between the zig-zag's instead of zero.

Because $F_n$ is piecewise linear between these fractions the
coefficients can be calculated directly using that $F_n(q/n)=q/n$ for
$\nu_B\le1/2$ and $\nu_B\ge1/2$ respectively.  The coefficients are
(see also Table 1)
\bea \label{q}  
   \begin{array}{lll}
     \alpha_n &=& 1\pm n \\
     \beta_n  &=& \mp q  
  \end{array}    \quad ,
\eea
in the intervals $[q/n; q/(n\pm1)]$.
Between these intervals $\nu_B \in [q/(n-1),(q+1)/(n+1)]$
they both vanish for $\nu_B\le1/2$.

For $1/2\le\nu_B\le1$ we find the very same coefficients just below and
above $q/n$
as in Eq. (\ref{q}). However, the range is different (see Table 1) and
between these intervals $F_n=2\nu_B-1$, i.e.  $\alpha_n=2,\beta_n=-1$.
The coefficients are also shown in Figs. 1+2 for $n=1,2,...,9$.
 
For larger filling fractions, $\nu_B\ge 1$,
the coefficients can be related as
\bea
   F_n(\nu_B) = F_n(\nu_B-1)+(2\nu_B - 1)  \,,
\eea 
i.e., the zig-zag behavior repeats
for all $F_n$ at $\nu_B=1+q/n$. 
Note that $F_0$ vanishes for $\nu_B\le 1$ but
increases linearly as $F_0=\nu_B-1$ with filling fraction $1\le\nu_B\le 2$.

\subsection{Particle-hole symmetry}

The coefficients $F_n$ and resulting energies can be related at
$\nu_B$ and $1-\nu_B$ by exploiting particle-hole symmetry.  Since
$M_BF_n(\nu_B)$ is the number of particles with a particle $n$ sites
away, $M_B(\nu_B- F_n(\nu_B))$ is the number of particles with a
hole $n$ sites away. The latter is by particle-hole symmetry also 
equal to the number of holes with a
particle $n$ sites away, which again is equal to $M_B(1-\nu_B-
F_n(1-\nu_B))$. Here $M_B(1-\nu_B)$ is the number of holes and
$M_BF_n(1-\nu_B))$ the number of holes with a hole $n$ sites
away. Equating the two numbers for the number of particle-hole
neighbors $n$ sites away gives
\bea
   F_n(\nu_B) = F_n(1-\nu_B) + \, 2\nu_B - 1  \,.
\eea 
By insertion in Eq. (\ref{E}) we obtain the relation 
\bea \label{Esym}
   E(\nu_B) = E(1-\nu_B) + \, (2\nu_B - 1)E(1) 
\eea
Here $E(1)=\sum_{n\ge1}U_n$ is the energy for $N=M_B$.

A similar relation has been derived for the ground state energy of the
FQHE also exploiting particle-hole symmetry \cite{MacDonald}
\bea \label{EsymF}
   E(\nu) = E(1-\nu) + \, (2\nu - 1)E(1) \,,
\eea
Although the two relations appear identical, they are not because
$\nu$ and $\nu_B$ are different. One obvious cause for this difference
is the neglect of $T$ in the EBHM energy of Eq. (\ref{Esym}). 
For $N\gg M_B$ the tight-binding approximation is poor as many states
overlap. Furthermore, the EBHM
energy does not take the Laughlin wave function into account when the
energy is calculated. This becomes increasingly important for increasing
filling fraction $N\to M$ because the Laughlin wave function contains
$N$ powers of $z$ whereas the EBHM wave function only has $M_B=M-N$ powers
of $z$. For these reasons the EBHM ground state energies is not 
applicable to the FQHE for large filling fractions. 

\subsection{Chemical potential gaps and superfluid states}

The chemical potential at zero temperature is now
\bea
  \mu = \left( \frac{dE}{dN}\right)_{M_B} = \sum_n \alpha_n U_n \,.
\eea
The chemical potential is therefore also constant between the 
fractions and exhibit plateaus as shown in Figs. 3+4.
At the fractions the chemical
potential is discontinuous and therefore an insulator -
but a conductor between the fractions.
A chemical potential gap appears in the spectrum at each fraction $q/n$,
\bea
   \Delta\mu(q/n)&=& \mu(q/n)_+-\mu_(q/n)_- \,,
\eea
where $\pm$ refers to the right/left side of the incompressible fraction
$q/n$. The gap
can be calculated from the expression (\ref{q}) for $\alpha_n$.
For $1/2<\nu_B<1$ we find
\bea
   \Delta\mu(q/n)&=& n(U_{n-1}-2U_n+U_{n+1}) \nonumber \\
   &\simeq&  n\left(\frac{d^2U_n}{dn^2}\right) \,.
\eea
Note that the convexity condition of Eq. (\ref{convex}) insures that the
gap is positive and the system therefore stable.

For $\nu_B<1/2$ we obtain
\bea
   \Delta\mu(q/n)&=& (n-2)U_{n-1}-2U_n+(n+2)U_{n+1} \nonumber\\
 &\simeq& n\left(\frac{d^2U_n}{dn^2}\right)-4\left(\frac{dU_n}{dn}\right)
  \,.
\eea
Again the convexity condition and that $dU_n/dn<0 $ insures that the gap is
positive.

If $q$ and $n$ have a common divisor as, e.g. 4/6=2/3,
then the gaps add up. The convexity condition may therefore be unnecessarily
strict for such fractions.

If the couplings decrease as a power law $U_n\propto n^{-\gamma}$,
then $\Delta\mu(q/n)=\gamma(\gamma+1)U_n/n$, and the gaps
decrease faster than $1/n$. Numerical calculations of a finite number
of electrons on a sphere find FQHE gaps that decrease slightly faster
than $1/n$ \cite{Morf}. The couplings $U_n\propto \log(M_B/n)$ leads to gaps
scaling as $\sim1/n$. The Coulombic interactions and wave functions localized
in $z$ would lead to $U_n\simeq e^2/l(n+const.)$ as discussed above.
The interaction strength may, however, be weakened by layer thickness and
effects of disorder \cite{Wan}.
 The
corresponding gaps would thus scale as $\sim e/l\, n^2$ at large $n$.

In the FQHE problem the repulsion between electrons is canceled on
average by a background field from the positively charged
solid. This mean field is, however, inert and can be ignored in
calculated the pressure and chemical potential of the quantum fluid.
Likewise, the optical lattice acts as an inert background of the
atomic gas systems.

\subsection{Pressure and incompressible states}

The pressure $P=-(dE/dV)=(\mu N-E)/V$ at zero temperature can be calculated
by  noting that the volume is proportional to the number of states $M_B$
\cite{area}. We shall simply set the volume per state to unity whereby
\bea
  P = -\left( \frac{dE}{dM_B}\right)_N = \mu \nu_B -\frac{E}{M_B} 
    = - \sum_n \beta_nU_n \,.
\eea
The pressure is therefore constant, i.e. compressible, between the 
fractions. If the couplings  $U_n$ are convex we observe that 
the pressure 
jumps to a higher pressure and 
the system is therefore an incompressible quantum fluid
at those fractions as shown in Figs. 3+4.
If the couplings $U_n$ are not convex at
$n$, the pressures may decrease at $\nu_B=q/n_+$ which would require a Maxwell
construction.

The pressure discontinuities
\bea
   \Delta P(q/n)&=& P(q/n_+)-P(q/n_-) \,,
\eea
can likewise be calculated by inserting the $\beta_n$ of Eq. (\ref{q}) in
the pressure. For $\nu_B<1/2$ we find 
\bea \label{Pq}
   \Delta P(q/n)
   &=& q(U_{n-1}-2U_n+U_{n+1}) \nonumber\\
   &\simeq& q\left(\frac{d^2U_n}{dn^2}\right) \,,
\eea
assuming that $U_n$ is a smooth function of $n$.
For $1/2<\nu_B<1$ we find 
\bea
   \Delta P(q/n)
   &=& (q-1)U_{n-1}-2qU_n+(q+1)U_{n+1} \,.
\eea

\begin{figure}
\begin{center}
\psfig{file=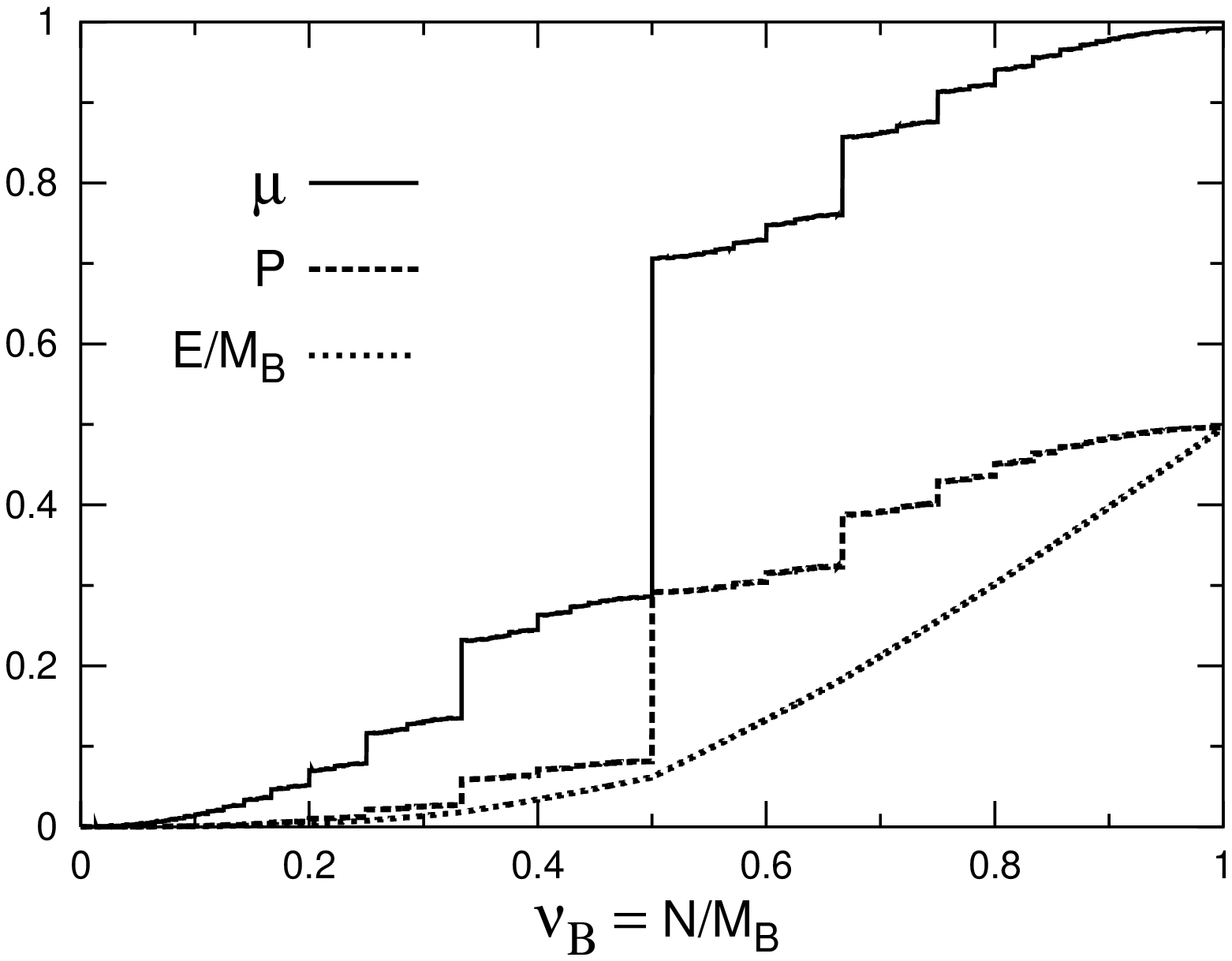,height=8.0cm,angle=-90}
\vspace{.2cm}
\begin{caption}
{The energy, chemical potential and pressure vs. filling fraction 
$\nu_B$. A short range two-body interaction is chosen with
neighbor couplings $U_n=3/(\pi n)^2$.
}
\end{caption}
\end{center}
\label{f3}
\end{figure}

\begin{figure}
\begin{center}
\psfig{file=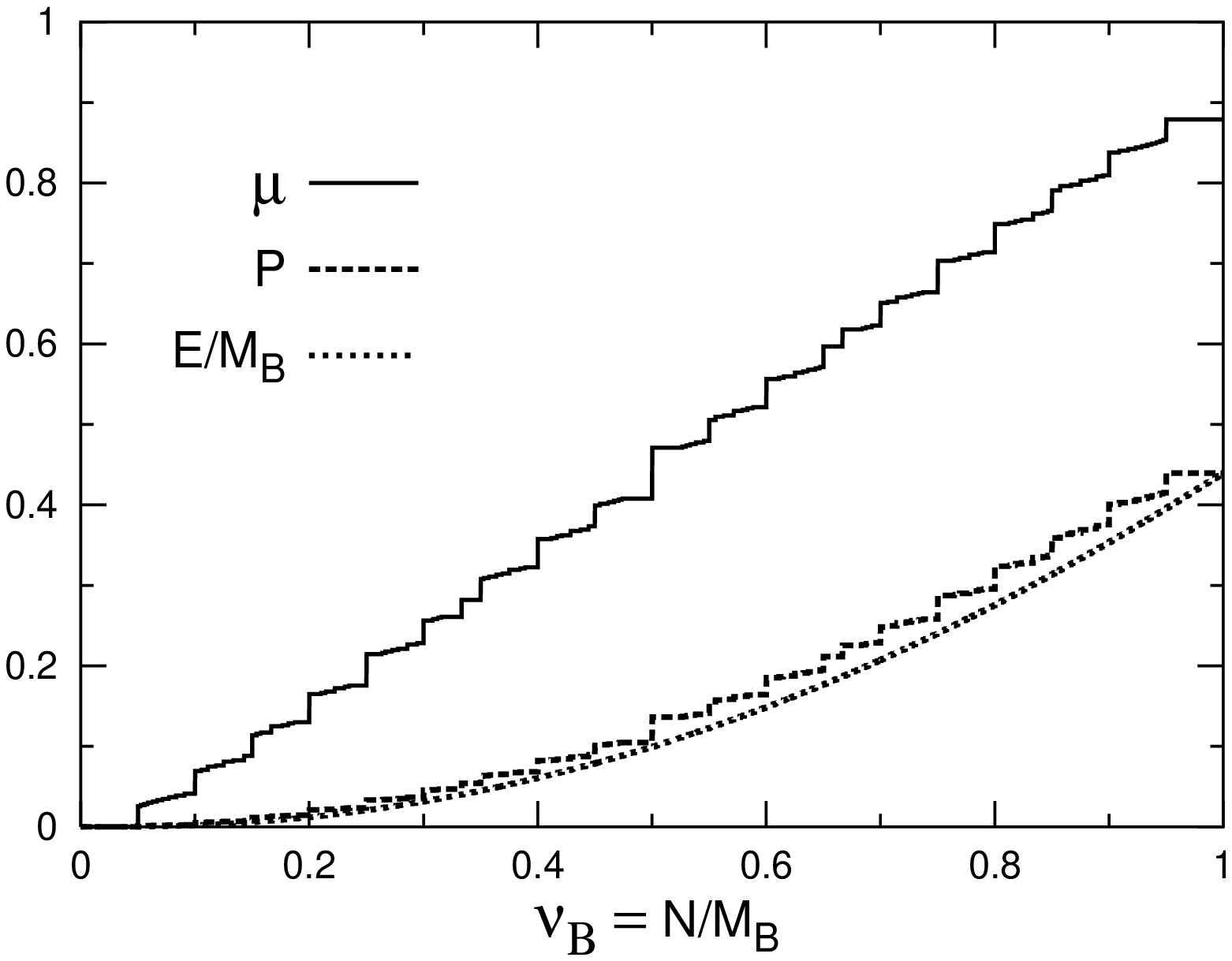,height=8.0cm,angle=-90}
\vspace{.2cm}
\begin{caption}
{As Fig. 3 but  with  
neighbor couplings $U_n=\log(M_B/n)/2M_B$ for $n\le M_B=20$, which
simulates a longer range two-body interaction.
}
\end{caption}
\end{center}
\label{f4}
\end{figure}

\subsection{Gap dependence on two-body range}

In Figs. 3+4 the energy, pressure and chemical potential are shown as
function of filling fraction $\nu_B$ for two representative set of
couplings.

The first assumes a short range (delta function) 
two-body interaction such that the couplings are $U_n=3/(\pi n)^2$
as discussed above. These couplings decrease rapidly with neighbor
distance $n$ and therefore the chemical potential gaps and pressure
discontinuities are largest at the fractions $q/n$ with small $n$,
e.g. $q/n=1/2$, 1/3, 2/3, 1/4, 3/4, etc.

In Fig. 4 a longer range two-body interaction is assumed such that the
neighbor coupling decrease more slowly with neighbor distance $n$.
The specific choice is $U_n=\log(M_B/n)/2M_B$ for $n\le M_B=20$ and
vanishing coupling for larger $n$. A logarithmic coupling is obtained
by averaging a two-body Coulomb potential over radial variables
preserving a phase difference $\phi_1-\phi_2=2\pi n/M_B$.  The rather
slow decrease of $U_n$ with $n$ has the effect that the gaps and
pressure discontinuities only decrease as $\sim 1/n$. The long range
interaction approaches the mean field limit $V=V_0$ where
$E/M_B=\nu_B^2/2=P$ and $\mu=\nu_B$.

\section{Incompressible states at fractional numbers}

The incompressible states found for the system of bosons above can 
as described in section II be
translated to the FQHE states for fermions if we assume that the
hopping term $T$ in the full Hamiltonian is small, or if
it does not destroy the incompressible state by making a Mott transition. 

The incompressible states in the EBHM appear at fractions $\nu_B=q/n$,
where $n$ and $q$ are integers. However, when applied to the FQHE, it is
important first to separate out the Laughling wavefunction with
maximal $m$ as mentioned in section II. For this reason values
$\nu_B<1/2$ do not come into play in the FQHE because the preferred
ground state wavefunction then includes a Laughlin wavefunction with a
$m$-value increased by 2.
 
According to Eq. (\ref{nu}) the Bose fractions correspond to FQHE
filling fractions
\bea
   \nu = \frac{\nu_B}{1+m\nu_B}
    = \frac{q}{n+mq} \,,
\eea
where $m=1,3,5,...$ is the Laughlin number. 
Most of the leading fractions are shown in Table II and 
will now be compared to the FQHE fractions
found in FQHE experiments \cite{Tsui,Pan}.

\begin{table}
\caption{\label{tab:table4} The leading incompressible filling fractions 
for small $n$ and $q$ for $m=1$ (see text). Observed FQHE fractions are 
given in bold face.}
\begin{ruledtabular}
\begin{tabular}{||r|l|l||}
 n & $q$ & $\nu=\nu_B/(1+m\nu_B)=q/(n+mq)$ \\
\hline
  1 & 1,2,3,4,5,6,... & 1/2, {\bf 2/3}, 3/4, {\bf 4/5}, 5/6, {\bf 6/7},...\\
  2 & 1,3,5,7,... & {\bf 1/3, 3/5, 5/7}, 7/9,... \\
  3 & 2,4,5,... & {\bf 2/5, 4/7, (5/8)},... \\
  4 & 3,5,7,... & {\bf 3/7, 5/9, 7/11)},... \\
  5 & 3,4,6,7,.. & {\bf 3/8, 4/9, 6/11}, 7/12, ..\\
  6 & 5,7,11,... & {\bf 5/11, 7/13}, 11/17,... \\
  7 & 4,5,6,8,... &  {\bf 4/11, (5/12), 6/13, 8/15}, ... \\
  8 & 5,7,9,11,... & {\bf 5/13, 7/15, 9/17}, 11/19,... \\
  9 & 5,7,8,10,11,.. & {\bf (5/14)}, 7/16, {\bf 8/17, 10/19}, 11/20,... \\
  10& 7,9,11,13,.. &  {\bf (7/17), 9/19, 11/21}, 13/23,... \\
  11& 6,7,... & {\bf 6/17}, 7/18,... \\ 
\end{tabular}
\end{ruledtabular}
\end{table}

\subsection{The Laughlin and Jain sequences}

The incompressible states at $\nu_B=1/2$ gives the
Laughlin fractions 1, 1/3, 1/5, 1/7, etc. for $m=1,3,5,7$, ... respectively.
Next to these the most important set of fractions is the Jain sequence
\bea \label{Jain}
   \nu = \frac{n}{2n\pm1} \,.
\eea
These appear for $m=1$.
Inserting the EBHM fractions 
$\nu_B=1-1/n$, $n=1,2,3,...$ in Eq. (\ref{nu}) gives the positive
(+) part of the principal FQHE fractions,
i.e. 1/3, 2/5, 3/7, etc.
There are analogous incompressible states
at fillings $\nu_B=1+1/n$. Inserting these in Eq. (\ref{nu})
they lead to the other (-) half of the principal fractions of Eq.
(\ref{Jain}), i.e. 2/3, 3/5, 4/7, etc. 

The Jain fractions will be leading for two reasons. Firstly, the pressure
discontinuity of Eq. (\ref{Pq}) increase with $q$ and $q=n-1$ is maximal
for these fractions for a given $n$. Secondly, in the region above 
$\nu_B=1-1/(n-1)$, the fraction $\nu_B=1-1/n$
will be the incompressible fraction with the smallest
$n$. As the pressure
discontinuity of Eq. (\ref{Pq}) also decreases with $n$ it will therefore
be the largest in this region.

\subsection{$\nu<1/3$}

It was discussed in section II that for fractions $\nu<1/m$,
$m=1,3,5,..$ the Laughlin wave function with $m$ should be
employed. For example, for $1/5<\nu<1/3$ (corresponding to
$1/4<\nu_B<1/2$ for $m=1$) the Laughlin wave function with $m=3$
should be employed instead with a larger $\nu_B$ such that
$\nu=\nu_B/(3\nu_B+1)$.

For $m=3$
the commensurable fillings $\nu_B=n$, $n=2,3,4,5,...,$ therefore lead to the
fractions 2/7, 3/10, 4/13, 5/16, etc. These first three 
are clearly identified in the
FQHE spectrum \cite{Pan}. 
The next leading fractions $\nu_B=3/2$, 5/2, 7/2, etc. gives
$\nu=2/9$, 5/17, 7/23, etc., respectively. The two first are also 
observed in the FQHE spectrum. The large $\nu_B$ become gradually more 
difficult to observe because the corresponding FQHE fraction approaches
1/3 and is therefore quenched by the Laughlin fraction $\nu=1/3$.

\subsection{$1/3<\nu<1/2$}

Secondary incompressible states also appear in the EBHM at the fractions 
$q/n$, $q=n-2,n-3,...$, etc. As their gap size decrease with $n$ 
their importance are ordered likewise (see Table II). 
The leading secondary fractions (not including the Laughlin
and Jain series) are therefore:
$\nu_B=3/5$,
4/7, 5/7, 5/8, 5/9, 7/9, 7/10, 6/11, etc.  These EBHM fractions
correspond to the FQHE fractions
$\nu=3/8$, 4/11, 5/12, 5/13, 5/14, 7/16, 7/17, 6/17, etc. (see also table II).
Remarkably, four of
these fractions are clearly identified in recent high resolution
FQHE experiments \cite{Pan}. The
fraction at 5/12 is compatible with the
structure reported between the principle fractions $\nu=2/5$ and 3/7.
The fractions 5/14 and 7/17 are compatible with structures
between 4/11 and 6/17. The remaining fraction 7/16 is
squeezed between the two Jain fractions 3/7 and 4/9 and thus
difficult to observe.

Incompressible states with fractions $q/n$ with larger $n$ will be
weaker. Except for the Jain fractions, they will also be placed between
stronger incompressible states, i.e. fractions with smaller $n$.
Therefore they will be increasingly difficult to identify experimentally.
The incompressibilities and chemical potential gaps become smaller
and eventually disappear with increasing $n$. 
Effects of impurities, finite temperature and experimental resolution
therefore exclude experimental observation of the
incompressible states at large $n$.

\subsection{$\nu=1/2$}

The EBHM is apparently successful in producing all the incompressible
states at the principle and secondary FQHE fractions including those
found in recent experiments.  However, it clearly fails for $m=1$ at half
filling or $\nu_B=1$ where it also predicts an incompressible state in
contradiction with FQHE experiments and theoretical calculations
\cite{Halperin}.  In the Bose-Hubbard model the state at $N=M_B$
becomes compressible for a sufficiently strong hopping term.  It will
be discussed in the next section that the hopping term $T$ may cause a
transition from a Mott insulator to a superconductor, i.e. changes the
state at $\nu=1/2$ to a compressible one.

\subsection{$\nu>1/2$}

For $\nu_B\ge 1$ the EBHM predicts a number of incompressible states. The
leading ones are shown in Table II. Experimentally, they are increasingly
difficult to observe as $\nu\to1$ due to limited resolution. The set of
states at
$\nu_B=2-1/n=5/3$, 7/4, 9/5, etc., correspond to FQHE states
$\nu=5/8$, 7/11, 9/14, respectively. 
The 7/11 is clearly observed whereas a structure is
found at 5/8 and possibly also at 9/14.

However, as mentioned above the ground state energies of the EBHM
become less reliable with increasing filling fraction as a model for
the FQHE.  When $\nu_B>1$ it is therefore a better approximation to
use the particle-hole symmetry of the FQHE, Eq. (\ref{EsymF}) by
inserting the EBHM energies for $\nu_B<1$ as an approximation for the
FQHE ground state energies.  It follows that the incompressible states
for $\nu$ also exist for $1-\nu$.

\section{Mott-Hubbard transitions}

The term $T$ contains a number of hopping terms between $n$'th
neighbors. It is therefore instructive to compare to the one
dimensional Bose-Hubbard model (Eq. (\ref{BH}) which contains nearest
neighbor hopping $t_1$ but only on-site interactions, i.e. only $U_0$
is non-vanishing. In the strongly interacting limit ($T\ll H_0$) it is
a Mott insulator for integer filling factors \cite{Fischer,Zwerger} as
also found above for the EBHM.  Increasing the hopping term a Mott
transition takes place at $t_1\simeq U_0/1.92$ in one dimension and
the system becomes a superfluid.
 
It is known from studies of the FQHE state at half filling (or
$N/M_B=1$) \cite{Halperin} that the state is compressible.  We should
therefore expect that by including $T$ the incompressible state at
$N=M_B$ undergoes a Mott transition.

Adding terms $U_n\ne 0$ but keeping $T=0$, creates new Mott
insulator states at the corresponding fractional filling factors as
described above.  A crucial question is therefore whether 
these Mott insulator states survive as the hopping term $T$ is gradually
turned on. 
Studies of the Bose-Hubbard model including next neighbor interactions
show that the Mott insulator state at half filling 
or charge density wave disappear before that at unity filling as the
nearest neighbor hopping $t_1$ is increased. $T$ is, however, very different
from $t_1$.

A closer look at the ground states for fractional and integer filling
fractions reveals a difference in the overlap between the 
degenerate ground states. The ground state for $N=M_B-1$
\bea 
  .....\bullet\bullet\bullet\bullet\bullet\bullet\bullet\circ\bullet\bullet
  \bullet\bullet\bullet\bullet\bullet\bullet .....
    \nonumber
\eea
is $M_B$-fold degenerate and the hopping term $T$ couples them all.
As a consequence the energy can change considerably. 
It is the far-neighbor hopping terms in $T$ that can be responsible for
``destabilizing'' the integer filling fraction states. Only the nearest
neighbor hopping $t_1$ is present in the Bose-Hubbard model.

In comparison the states near fractional filling, e.g.
$N=2M_B/3-1$
\bea 
  ....|\bullet\bullet\circ|\bullet\bullet\circ|\bullet\bullet\circ|
  \bullet\circ\bullet\circ|
  \bullet\bullet\circ|\bullet\bullet\circ|\bullet\bullet\circ| 
  \bullet\bullet\circ| .....
    \nonumber
\eea
this state only couple to two of the $N$ degenerate ground states,
namely those where the extra hole hops one site to one side, 
resulting in a modest energy corrections.

Another difference appears when comparing the corrections to the
ground state energies just above and below $N/M_B=2/3$ with those
at $N=M_B$. In the former case the correction from including hopping
terms from $T$ is the same for $N=2M_B/3\pm1$. However, the leading
hopping term correction
for $N=M_B\pm1$ differs because it vanishes when one particle is added
to commensurate filling. 

It should be noted that only the odd commensurate states $\nu_B=1,3,5,...$
should undergo a Mott transition and only for $m=1$ in order to fully
agree with the FQHE.  It has be argued that the far hopping terms are
important for commensurate states but at present the EBHM cannot
explain why only the odd commensurate states for $m=1$ vanish for the FQHE. 
It may
be related to the difference in energy between the energies of the
EBHM and FQHE, Eqs. (19) and (20) respectively. The Laughlin
wavefunction may affect energies of the FQHE especially at large
filling $\nu_B$ and $m$.

The EBHM is at most an approximation to the
FQHE since the localized states are only a good approximation when
the hopping terms are small. Furthermore, a bosonization
procedure was applied without including the Laughlin wave function,
when energies were calculated within the EBHM.  These approximations
should be investigated further.

 The effects of far-neighbor hopping is a complicated yet important
problem. Experimentally, it may be studied by quantum gases in optical
lattices where the neighbor couplings and hopping can be controlled by
tuning the lattice spacing and height, interaction strengths and
particle density. Both the neighbor couplings $U_n$ and the hopping
coefficients $t_n$ decrease exponentially with the tunneling factor
$\sim\exp(-2n\sqrt{V_0/E_R})$, where $V_0$ is the optical lattice
amplitude and $E_R$ the recoil energy \cite{Zwerger}.
In addition, however,
the couplings $U_n$ also scale with the scattering length which allows us
to enhance the couplings near a Feshbach resonance and thus reach the 
strongly interacting limit. If absorption images can be obtained from
such a one dimensional optical lattice then
Bragg peaks should be observed displaying
fractional reciprocal lattice vectors due to the periodicity at
fractional filling.

\section{Summary and outlook}

The Bose-Hubbard model extended by including nearest and far neighbor
interactions was related to the fractional quantum Hall effect
problem. The ground state is calculated for the extended Bose-Hubbard
model in the strongly repulsive limit (weak hopping). Incompressible
Mott insulator states were found at filling fractions compatible with
the FQHE filling fractions of the lowest Landau level. The
incompressible states with the largest gaps are the Laughling and Jain
fractions.  Secondary incompressible states occur for filling
fractions 3/8, 4/11, 5/13, 6/17, 2/7, 3/10, 4/13, 5/16, 2/9, 5/17.
Such fractions have been observed in recent FQHE experiments. 
Weaker states at 9/14,
5/12, 5/14, etc. are also predicted and indications of structures at
the fractions are reported in recent high resolution FQHE experiments.
Furthermore, incompressible states at fractions with even
larger denominators are also predicted; the next candidates would
be 7/16, 7/17, 7/23, etc. However, these states are increasingly
difficult to observe and is a challenge to find experimentally.

The EBHM provides an interesting spectrum of incompressible state at
fractional fillings. Whether these states survive or undergoes a Mott
insulator transition to a superfluid as the hopping terms are included
is an important question. It has been indicated that far hopping terms
are important for commensurate states $\nu_B=1,2,3,...$ but it is not
clear why only the odd ones should undergo a Mott transition. The EBHM is at
most an approximation to the FQHE since the localized states are only
a good approximation when the hopping terms are small. Furthermore, a
bosonization procedure was applied without including the Laughlin wave
function, when energies were calculated within the EBHM.  These
approximations should be investigated further.

In spite of its limitations as a model for the FQHE, the EBHM does
provide a ground state wave function in the strongly interacting limit
from which incompressible fractions, gaps and incompressibilities can
be calculated. Corrections from hopping terms may be estimated
perturbatively using this wave function as a starting point.  Therefore,
the EBHM may provide a useful basis for also calculating a number of
transport coefficients such as the longitudinal and transverse Hall
resistances as well as their low temperature dependence.

The Mott transitions at fractional fillings between insulator and
superfluid phases may generally be studied by experiments with quantum
gases in optical lattices. 
By lowering the lattice heights the overlap between nearest and
next-nearest neighbors increase and therefore also $U_1$, $U_2$,...,
and the corresponding hopping terms \cite{Mazzarella}.  In optical
lattices the neighbor couplings and hopping can be controlled by
tuning the lattice spacing and height as well as the interaction
strengths between atoms and the particle density.  Furthermore, Bose
and Fermi atoms can be mixed resulting in a multitude of Mott, charge
and spin density wave, superfluid and other phases \cite{Illuminati,Wang}.

In high temperature superconductivity (HTc) ordered charge density
waves are found in the spin glass phase. These are ordered
1D stripe and 2D checkerboard phases with magic hole doping fractions
at $x=1/16, 3/32, 1/8, 3/16$ \cite{Komiya}.  These phases and their
competition with HTc are believed to be described by (extended) 2D
Hubbard Hamiltonians and therefore such models are actively
investigated. The ordering, Mott insulator phases and transitions in
lattices with fermions and/or bosons is an important topic in many
fields.



\appendix

\section{Localized states on a circle}

To motivate the basis of localized wave functions in Eq. (\ref{Phi}) 
we discuss the model of Dyakonov \cite{Dyakonov}, where
the FQHE problem is approximated by putting the states on a circle.

The idea behind the circle approximation is to replace the coordinates
$z=r e^{i\phi}$
by their phase: $e^{i\phi}$, by approximating 
the radius by its average value: $\av{|z|^2}=\av{r^2}=k+1$,
in units of the magnetic length. 
The states now become: $\psi_k\simeq\exp(ik\phi)$,
$k=0,1,2,...,M_B-1$.
The Laughlin wave function is then
\bea
   \Psi_m= A \prod_{i<j}(e^{i\phi_i}-e^{i\phi_j})^{m} \,,
\eea
with $m=1,3,5,...$ and normalization $A^2=(m!/2\pi)^N/(mN)!$.

The Wannier-type localized functions analogous to (\ref{Phi}) are simply 
\bea \label{Phic}
  \Phi_j(\phi) &=&  \frac{1}{\sqrt{2\pi M_B}}
      \sum_{k=0}^{M_B-1} e^{ik(\phi-2\pi j/M_B)} \nonumber \\
    &=& \frac{1}{\sqrt{2\pi M_B}}\, 
        \frac{1-e^{iM_B\phi}}{1-e^{i(\phi-2\pi j/M_B)}} \,,
\eea
where $j=0,1,...,M_B-1$. These wave functions are localized in angle $\phi$
around $2\pi j/M_B$. They oscillate rapidly with frequency $1/M_B$ and
amplitude dropping off slowly as $j^{-1}$.
They are the same function 
(a phase times $\sin(M_Bx)/\sin(x)$ with $x=\phi/2-\pi j/M_B$)
of the shifted angle $(\phi-2\pi j/M_B)$.
The localization
sites are relative to each other and can all be shifted by an arbitrary angle.
A many-body wave function build of such localized states would therefore 
rather resemble a strongly correlated liquid than a crystal.

On the circle the $V_{ij,i'j'}$  are simply
\bea
  V_{ij,i'j'} &=& \int \Phi_i(\phi_1)\Phi_j(\phi_2) 
                  \Phi^*_{i'}(\phi_1)\Phi^*_{j'}(\phi_2) \nonumber \\
   && |\Psi_m(\phi_1,\phi_2,...)|^2  V(\phi_1-\phi_2)d\phi_1 d\phi_2   \,,
\eea
where now
the two-body potential $V(\phi_i-\phi_j)$ depends on the angular distance
instead of $|z_i-z_j|$ as in Eq. (\ref{HV}). 

The relative simple localized wave functions allows us to study the
overlap integrals in $V_{ij,i'j'}$.  The localized wave functions are
rapidly oscillations and generally out of phase except for the
coupling between two sites. As result the direct coupling $V_{ij,ij}$
and the exchange $V_{ij,ji}$ are the largest. We therefore expect that
the tight binding approximation of Eq. (\ref{H0}), which includes only
the direct and exchange interactions, is a good approximation. This is
especially the case for long range interactions which approaches the
mean field limit. In this limit
$V_{ij,i'j'}=\delta_{i,i'}\delta_{j,j'}\int V(\phi)d\phi$.

For interactions of range shorter than $\la 1/M_B$, the two-body
potentials can be approximated by a delta-function interaction:
$V(\phi)\simeq 2\pi V_0\delta(\phi)$.
The resulting $U_n$ for the delta function interaction are
\bea \label{Uphi}
  U_n = 2\pi V_0 \int_0^{2\pi} |\Phi_i(\phi)|^2|\Phi_{i+n}(\phi)|^2
   \, d\phi   \,.
\eea
Here, we have ignored the Laughlin
wavefunction but will discuss the effect of the correlation hole below.
Note that generally $\sum_n U_n =M_B\int V(\phi)d\phi$ 
and for the delta-function
interaction  $\sum_n U_n =2\pi M_BV_0$. Also $U_n=U_{M_B-n}$.

Inserting the wave functions of Eq. (\ref{Phic}) in (\ref{Uphi}) yields
$U_0=(2/3)M_BV_0$ and $U_n=M_BV_0[M_B\sin(\pi n/M_B)]^{-2}$ for $n\ge1$.
In the thermodynamic limit $M_B\to\infty$ the couplings scale as
$U_n=M_BV_0/(\pi n)^{2}$. For longer range interaction the couplings
$U_n$ decrease with $n$ at a slower rate. For a Coulomb-type potential
$V=e^2/|\phi_1-\phi_2|$ we obtain $U_n\sim M_B e^2/(n+const.)$, where the
constant is of order unity and reflects that $U_n$ is finite due to the 
correlation hole in the Laughlin wave function.

The delta-function approximation for the short range interaction
should only be employed for the bosonic part of the wave function. As a
consequence the direct and exchange interaction energies are the
same and leads to a factor two.  The delta function interaction should
not be employed for the fermionic part of the wave function.  The
exchange part would then cancel the direct part and lead to vanishing
interaction energy as it should for an antisymmetric wave function.

For a longer range potential the exchange and direct parts do not cancel.
Also the couplings $U_n$ will decrease slower with $n$. 
Such an example $U_n\propto log(M_B/n)$ is shown in Fig. 4.

Returning to the Coulomb potential in coordinate space
$V=e^2/|z_1-z_2|$ and the localized wave functions of Eq. (\ref{Phi}),
the couplings of Eq. (\ref{J}) with (\ref{Vij}) become $U_n\sim
(e^2/\av{r})\log(M_B/n)$. These wave functions are, however, only
localized in their phases. Therefore the 2-D integration over both
radius and angle is different than the 1D integration over angle on
the circle. The result is a $\log(M_B/n)$ in stead of $1/n$
respectively.
  
It may be advantageous to try to construct a basis of wave functions
that are localized both radially and in angle $\phi$, i.e. in
localized in $z=re^{i\phi}$.  For a Coulomb potential the couplings
will also be approximately Coulombic $U_n\simeq
e^2/l(n+const.)$. Here, the constant takes into account that the
coupling $U_0$ is finite according to definitions in Eq. (\ref{Vij}).

\section{Tables of $F_n$ coefficients}

The first coefficients $F_n=\alpha_n\nu_B+\beta_n$
are listed in Tables III+IV for filling fractions $\nu_B\le1$ and $n\le7$.

\begin{table}
\caption{\label{tab:table2} The coefficients in $F_n=\alpha_n\nu_B+\beta_n$
for $n\le4$.}
\begin{ruledtabular}
\begin{tabular}{||r|r|r|l||}
distance n& $\alpha_n$ & $\beta_n$ & range of $\nu_B$ \\
\hline
 0 & 0 & 0  & [  0 , 1 ] \\
   & 1 & -1 & [  1 , 2 ] \\
\hline
 1 & 0 & 0  & [ 0 , 1/2 ] \\
   & 2 & -1 & [ 1/2 ,  1 ] \\
\hline
 2 & 0  & 0 & [  0  , 1/3 ] \\
   & 3 & -1 & [ 1/3 , 1/2 ]\\
   & -1 & 1 & [ 1/2 , 2/3 ]\\
   & 2 & -1 & [ 2/3 ,  1  ]\\
\hline
 3 & 0 & 0  & [  0  , 1/4 ] \\
   & 4 & -2 & [ 1/4 , 1/3 ] \\
   & -2 & 1 & [ 1/3 , 1/2 ]\\
   & 4 & -2 & [ 1/2 , 2/3 ]\\
   & -2 & 2 & [ 2/3 , 3/4 ]\\
   & 2 & -1 & [ 3/4 ,  1 ]\\
\hline
 4 & 0 & 0  & [ 0 , 1/5 ] \\
   & 5 & -1 & [ 1/5 , 1/4 ] \\
   & -3 & 1 & [ 1/4 , 1/3 ] \\
   & 0 & 0  & [ 1/3 , 2/5 ]\\
   & 5 & -2 & [ 2/5 , 1/2 ]\\
   & -3 & 2 & [ 1/2 , 3/5 ]\\
   & 2 & -1 & [ 3/5 , 2/3 ]\\
   & 5 & -3 & [ 2/3 , 3/4 ]\\
   & -3 & 3 & [ 3/4 , 4/5 ]\\
   & 2 & -1 & [ 4/5 , 1 ]\\
\end{tabular}
\end{ruledtabular}
\end{table}

\begin{table}
\caption{\label{tab:table3} The coefficients in $F_n=\alpha_n\nu_B+\beta_n$
for $n=5,6,7$ in the interval of fractions $1/2\le\nu_B\le1$.}
\begin{ruledtabular}
\begin{tabular}{||r|r|r|l||}
 n& $\alpha_n$ & $\beta_n$ & range of $\nu_B$ \\
\hline
 5 & 6 & -3 & [ 1/2 , 3/5 ]\\
   & -4 & 3 & [ 3/5 , 2/3 ]\\
   & 2 & -1 & [ 2/3 , 3/4 ]\\
   & 6 & -4 & [ 3/4 , 4/5 ]\\
   & -4 & 4 & [ 4/5 , 5/5 ]\\
   & 2 & -1 & [ 5/6 , 1 ]\\
\hline
 6 & -5 & 3 & [ 1/2 , 4/7 ]\\
   & 2 & -1 & [ 4/7 , 3/5 ]\\
   & 7 & -4 & [ 3/5 , 2/3 ]\\
   & -5 & 4 & [ 2/3 , 5/7 ]\\
   & 2 & -1 & [ 5/7 , 4/5 ]\\
   & 7 & -5 & [ 4/5 , 5/6 ]\\
   & -5 & 5 & [ 5/6 , 6/7 ]\\
   & 2 & -1 & [ 6/7 , 1 ]\\
\hline
 7 & 8 & -4 & [ 1/2 , 4/7 ]\\
   & -6 & 4 & [ 4/7 , 5/8 ]\\
   & 2 & -1 & [ 5/8 , 2/3 ]\\
   & 8 & -5 & [ 2/3 , 5/7 ]\\
   & -6 & 5 & [ 5/7 , 3/4 ]\\
   & 2 & -1 & [ 3/4 , 5/6 ]\\
   & 8 & -6 & [ 5/6 , 6/7 ]\\
   & -6 & 6 & [ 6/7 , 7/8]\\
   & 2 & -1 & [ 7/8 , 1]\\
\end{tabular}
\end{ruledtabular}
\end{table}


\begin{thebibliography}{99}
\bibitem{area} For a sample area $S$ the number of available states in
the lowest Landau level is $M=S/2\pi l^2$ where $l=\sqrt{\hbar c/eB}$
is the magnetic length.
\end{thebibliography}
\end{document}